\documentclass[12pt]{article}
\usepackage{amsmath}
\usepackage{amssymb}
\usepackage{amsfonts}
\usepackage{graphics}
\usepackage{epsfig}
\usepackage{latexsym}
\usepackage{manfnt}

\begin{document}

\noindent

\hfill \qquad\qquad Modified, May 2012

\renewcommand{\theequation}{\arabic{section}.\arabic{equation}}
\thispagestyle{empty}
\vspace*{-1,5cm}
\noindent \vskip3.3cm

\begin{center}
{\Large\bf On generating functions of Higher Spin cubic interactions}

\bigskip
{\large Karapet Mkrtchyan}
\bigskip

{\small\it Yerevan Physics Institute\\ Alikhanian Br.
Str.
2, 0036 Yerevan, Armenia}\\
\smallskip
and\\
\smallskip
{\small\it Department of Physics\\ Erwin Schr\"odinger Stra\ss e \\
Technical University of Kaiserslautern, Postfach 3049}\\
{\small\it 67653
Kaiserslautern, Germany}\\
\medskip
{\small\tt karapet@yerphi.am}\footnote{From October, 2011 - Scuola Normale Superiore and INFN,
Piazza dei Cavalieri 7, 56126 Pisa, Italy, karapet.mkrtchyan@sns.it}
\end{center}\vspace{2cm}

\bigskip
\begin{center}
{\sc Abstract}
\end{center}
\quad
We present off-shell generating functions for all cubic interactions of totally symmetric massless Higher Spin gauge fields and discuss their properties.

\bigskip
\bigskip
\bigskip
\bigskip
\bigskip
\bigskip

\bigskip
\bigskip
\bigskip
\bigskip
\bigskip
\bigskip
\bigskip
\bigskip

\quad Based on talk, given at the conference ``14th International Conference on Symmetry Methods in Physics (SYMPHYS-14)'', 16-22 August 2010. Tsakhkadzor, Armenia.

\newpage

\section{Introduction and notations}

\quad

Despite the success of Vasiliev equations \cite{VasilievEqn}, consistent deformation of Higher Spin free Lagrangian \cite{Fronsdal:1978rb} to full nonlinear interacting theory for any spacetime dimensions is still an open task.

After Metsaev's light-cone classification of the cubic vertices for massive and massless higher spin fields \cite{Metsaev:2007rn}, recently the cubic vertices for totally symmetric massless fields were constructed and classified in the covariant approach \cite{Manvelyan:2010jr}, generalizing Noether procedure technic of \cite{Manvelyan:2010wp}. The results are in full agreement with Metsaev's classification. The interactions of \cite{Metsaev:2007rn}, \cite{Manvelyan:2010jr} are unique for given spins and number of derivatives, include all possibilities for parity preserving cubic interactions of higher spin fields in Minkowski space of any dimension greater or equal to four. It was shown in \cite{Sagnotti:2010at} that all of these vertices are realized in string theory. Then off-shell generating functions of cubic interactions for both reducible \cite{Fotopoulos:2010ay} and irreducible symmetric \cite{Manvelyan:2010je} higher spin fields became available. For symmetric fields, the results of \cite{Fotopoulos:2010ay} reproduced vertices, known from \cite{Bengtsson:1987jt}. This recent development provided new insight into earlier works \cite{Bengtsson:1983pd}-\cite{Metsaev:1993ap}. For more recent literature see \cite{Metsaev:2005ar}-\cite{Schlotterer:2010kk} and references therein. We present here general, but compact form for interactions between fields of irreducible Fronsdal setting, using the results of \cite{Manvelyan:2010je}.

To continue with this subject we introduce here briefly our notations (see for example \cite{Manvelyan:2008ks}). As usual, instead of symmetric tensors such as $h^{(s)}_{\mu_1\mu_2...\mu_s}(z)$, we use homogeneous polynomials in the vector $a^{\mu}$ of degree $s$ at the base point $z$
\begin{equation}
h^{(s)}(z;a) = \sum_{\mu_{i}}(\prod_{i=1}^{s} a^{\mu_{i}})h^{(s)}_{\mu_1\mu_2...\mu_s}(z) .
\end{equation}
Then we have for symmetrized gradient, trace and divergence \footnote{To distinguish easily between "a" and "z" spaces we introduce the notation $\nabla_{\mu}$ for space-time derivatives $\frac{\partial}{\partial z^{\mu}}$.}
$$
(a\nabla)h^{(s)}(z;a),\quad \frac{1}{s(s-1)}\Box_{a}h^{(s)}(z;a),\quad \frac{1}{s}(\nabla\partial_{a})h^{(s)}(z;a).
$$
where
$$
a\nabla=a^{\mu}\nabla_{\mu}, \quad \Box_a=\frac{\partial}{\partial a_{\mu}}\frac{\partial}{\partial a^{\mu}}, \quad \nabla\partial_{a}=\nabla_{\mu}\frac{\partial}{\partial a_{\mu}}
$$
and summation for repeating indices apply.

\section{Free Lagrangian for all higher spin gauge fields}

\setcounter{equation}{0}
\quad We introduce a generating function for HS gauge fields as
\begin {equation}
\Phi(z;a) = \sum_{s=0}^{\infty} \frac{1}{s!}h^{(s)}(z;a) \label{sf}
\end{equation}
where we assume that all terms in the generating function for higher spin gauge fields (\ref{sf}) have the same scaling dimension.

Lowest order linearized gauge transformation for this field reads as
\begin{eqnarray}
\delta^{0}_{\Lambda}\Phi(z;a) = (a\nabla)\Lambda(z;a)\label{gv},\\
\delta^{0}_{\Lambda}D_{a}\Phi(z;a) = \Box \Lambda(z;a),\\
\delta^{0}_{\Lambda}\Box_{a}\Phi(z;a) = 2(\nabla\partial_{a})\Lambda(z;a).
\end{eqnarray}
where
\begin{eqnarray}
\Lambda(z;a)=\sum_{s=1}^{\infty} \frac{1}{(s-1)!}\epsilon^{(s-1)}(z;a)\label{gp},
\end{eqnarray}
is the generating function of the gauge parameters.
Fronsdal's constraint for the gauge parameter reads as
\begin{eqnarray}
\Box_{a}\Lambda(z;a)=0\label{FC},
\end{eqnarray}
The Fronsdal constraint on the gauge field reads in these notations
\begin{eqnarray}
\Box_{a}^{2}\Phi(z;a)=0\label{FC2},
\end{eqnarray}
We introduced the ``de Donder'' operator
\begin{equation}
D_{a_i} = (\partial_{a_i}\nabla_i)-\frac{1}{2}(a_i\nabla_i)\Box_{a_i} \label{D}
\end{equation}

Now we can write the free Lagrangian for all gauge fields of any spin in the following form
\begin{eqnarray}\label{free}
\mathcal{L}^{\emph{free}}(\Phi(z))&=&\frac{\kappa}{2} \exp[\lambda^{2}\partial_{a_{1}}\partial_{a_{2}}]\int_{z_{1}z_{2}} \delta (z_{1}-z) \delta(z_{2}-z)\nonumber\\
                            &&\{(\nabla_{1}\nabla_{2})-\lambda^{2}D_{a_1}D_{a_2}-\frac{\lambda^{4}}{4}
                            (\nabla_{1}\nabla_{2})\Box_{a_1}\Box_{a_2}\}
                            \Phi(z_{1};a_{1})\Phi(z_{2};a_{2})\mid _{a_{1}=a_{2}=0}\qquad\quad
\end{eqnarray}
where $\lambda^{2}$ compensates the scaling dimension of the operator in the exponent. In the free Lagrangian (\ref{free}) there is no mixing between gauge fields of different spin. Hence this expression reproduces Fronsdal's Lagrangians for gauge fields with any spin.
The parameter $\kappa$ is a constant which makes the action dimensionless.

\section{Cubic Interactions}
\setcounter{equation}{0}
\quad We are going to present a compact form of all HS gauge field interactions derived in covariant form in \cite{Manvelyan:2010jr}.
First we rewrite the leading term of a general cubic interaction of higher spin gauge fields with any spins $s_{1},s_{2},s_{3}$
\footnote{$\nabla_{ij}=\nabla_{i}-\nabla_{j}$, $\nabla_{2}\partial_{a}=\frac{\partial}{\partial a^{\mu}}\nabla_{2}^{\mu}$, $\partial_{a}\partial_{b}=\frac{\partial}{\partial a^{\mu}}\frac{\partial}{\partial b_{\mu}}$ and analogously for others.}
\begin{eqnarray}
&&\mathcal{L}^{leading}_{(1)}(h^{(s_{1})}(z),h^{(s_{2})}(z),h^{(s_{3})}(z))\nonumber\\
&&=\int_{z_{1},z_{2},z_{3}} \delta(z-z_{1})\delta(z-z_{2})\delta(z-z_{3})
(\nabla_{12}\partial_{c})^{s_{3}-n}(\nabla_{23}\partial_{a})^{s_{1}-n}
(\nabla_{31}\partial_{b})^{s_{2}-n}\nonumber\\
&&\left[(\partial_{a}\partial_{b})(\nabla_{12}\partial_{c})+
(\partial_{b}\partial_{c})(\nabla_{23}\partial_{a})+(\partial_{c}\partial_{a})
(\nabla_{31}\partial_{b})\right]^n
h^{(s_{1})}(a;z_{1})h^{(s_{2})}(b;z_{2})h^{(s_{3})}(c;z_{3})\qquad\quad\label{leading}
\end{eqnarray}
where the number of derivatives is
\begin{eqnarray}
\Delta=s_{1}+s_{2}+s_{3}-2n,\\
0\leq n\leq min(s_{1},s_{2},s_{3})
\end{eqnarray}
As we can see, the minimal and maximal possible numbers of derivatives are
\begin{eqnarray}
\Delta_{min}&=&s_{1}+s_{2}+s_{3}-2min(s_{1},s_{2},s_{3}),\\
\Delta_{max}&=&s_{1}+s_{2}+s_{3}.
\end{eqnarray}

These interactions trivialize only if we have two equal spin values and the third value is odd.
In that case we should have a multiplet of fields, with at least two charges, to couple to the odd spin field.
In the case of odd spin selfinteraction, the number of possible charges in the multiplet should be at least 3.

Now we can see that the following expression is a generating function for the leading term of all interactions of HS gauge fields.
\begin{eqnarray}
\mathcal {A}^{0}(\Phi(z)) = \int_{z_{1},z_{2},z_{3}} \delta(z-z_{1,2,3}) f(\hat W + \hat v) \times\Phi(z_1;a_{1})\Phi(z_2;a_{2})\Phi(z_3;a_{3})\mid_{a_{1}=a_{2}=a_{3}=0}\quad\label{exp1}
\end{eqnarray}
where $f$ is an arbitrary smooth function and
\begin{eqnarray}
&\hat W = \frac{\lambda^{2}}{2}[(\partial_{a_{1}}\partial_{a_{2}})(\partial_{a_{3}}\nabla_{12})+(\partial_{a_{2}}\partial_{a_{3}})(\partial_{a_{1}}\nabla_{23})
+(\partial_{a_{3}}\partial_{a_{1}})(\partial_{a_{2}}\nabla_{31})],\nonumber \\
&\hat v=\frac{1}{2}[(\partial_{a_{3}}\nabla_{12})+(\partial_{a_{1}}\nabla_{23})+(\partial_{a_{2}}\nabla_{31})]\label{W1},\\
&\int_{z_{1},z_{2},z_{3}} \delta(z-z_{1,2,3})=\int_{z_{1},z_{2},z_{3}} \delta(z-z_{1})\delta(z-z_{2})\delta(z-z_{3})
\end{eqnarray}
for brevity. Furthermore we will always assume this integration with delta functions, without writing it explicitly.
We assume that operator in the second row of (\ref{W1}) does not need any dimensionful constant multiplier.

Taking gauge variation of $\mathcal {A}^{00}$, and performing Neother procedure one can find generating functions for all other terms in the cubic Lagrangian.
We will make a shortcut, using the results of \cite{Manvelyan:2010je}. First we introduce new Grassmann-odd variables  $\eta_{a_{1}}, \bar{\eta}_{a_{1}}, \eta_{a_{2}}, \bar{\eta}_{a_{2}}, \eta_{a_{3}}, \bar{\eta}_{a_{3}}$.
Then we change expressions in the formula (\ref{exp1}) in a following way
\begin{eqnarray}
&&(\partial_{a_{i}}\partial_{a_{j}}) \rightarrow (\partial_{a_{i}}\partial_{a_{j}})+\frac{1}{4}\eta_{a_{i}}\bar{\eta}_{a_{j}}\Box_{a_{j}}
+\frac{1}{4}\eta_{a_{j}}\bar{\eta}_{a_{i}}\Box_{a_{i}},\\
&&(\partial_{a_{i}}\nabla_{jk}) \rightarrow (\partial_{a_{i}}\nabla_{jk})+\eta_{a_{i}}\bar{\eta}_{a_{j}}D_{a_{j}}-\eta_{a_{i}}\bar{\eta}_{a_{k}}D_{a_{k}}.
\end{eqnarray}
So we can write
\begin{eqnarray}\label{gf}
&\mathcal {A}(\Phi(z)) = \int d^6 \eta
f(\eta_{a_{1}}\bar{\eta}_{a_{1}}+\eta_{a_{2}}\bar{\eta}_{a_{2}}+\eta_{a_{3}}\bar{\eta}_{a_{3}}+W+\beta v)\nonumber\\
&\times\Phi(z_1;a_{1})\Phi(z_2;a_{2})\Phi(z_3;a_{3})\mid_{a_{1}=a_{2}=a_{3}=0}\qquad\quad\label{exp}
\end{eqnarray}
where
\begin{eqnarray}
&d^6 \eta= d\eta_{a_{1}}d\bar{\eta}_{a_{1}}d\eta_{a_{2}}d\bar{\eta}_{a_{2}}d\eta_{a_{3}}d\bar{\eta}_{a_{3}}\\
&W = \frac{1}{2}[\lambda^{2}(\partial_{a_{1}}\partial_{a_{2}}+\frac{1}{4}\eta_{a_{1}}\bar{\eta}_{a_{2}}\Box_{a_{2}}
+\frac{1}{4}\eta_{a_{2}}\bar{\eta}_{a_{1}}\Box_{a_{1}})]
[\partial_{a_{3}}\nabla_{12}+\eta_{a_{3}}\bar{\eta}_{a_{1}}D_{a_{1}}
-\eta_{a_{3}}\bar{\eta}_{a_{2}}D_{a_{2}}]\quad\quad\nonumber\\
&+\frac{1}{2}[\lambda^{2}(\partial_{a_{2}}\partial_{a_{3}}+\frac{1}{4}\eta_{a_{2}}\bar{\eta}_{a_{3}}\Box_{a_{3}}
+\frac{1}{4}\eta_{a_{3}}\bar{\eta}_{a_{2}}\Box_{a_{2}})]
[\partial_{a_{1}}\nabla_{23}+\eta_{a_{1}}\bar{\eta}_{a_{2}}D_{a_{2}}
-\eta_{a_{1}}\bar{\eta}_{a_{3}}D_{a_{3}}]\quad\nonumber\\
&+\frac{1}{2}[\lambda^{2}(\partial_{a_{3}}\partial_{a_{1}}+\frac{1}{4}\eta_{a_{3}}\bar{\eta}_{a_{1}}\Box_{a_{1}}
+\frac{1}{4}\eta_{a_{1}}\bar{\eta}_{a_{3}}\Box_{a_{3}})]
[\partial_{a_{2}}\nabla_{31}+\eta_{a_{2}}\bar{\eta}_{a_{3}}D_{a_{3}}
-\eta_{a_{2}}\bar{\eta}_{a_{1}}D_{a_{1}}]\label{W0}\quad\\
\break
&v=[\partial_{a_{3}}\nabla_{12}+\eta_{a_{3}}\bar{\eta}_{a_{1}}D_{a_{1}}
-\eta_{a_{3}}\bar{\eta}_{a_{2}}D_{a_{2}}]\quad\nonumber\\
&+[\partial_{a_{1}}\nabla_{23}+\eta_{a_{1}}\bar{\eta}_{a_{2}}D_{a_{2}}
-\eta_{a_{1}}\bar{\eta}_{a_{3}}D_{a_{3}}]\nonumber\\
&+[\partial_{a_{2}}\nabla_{31}+\eta_{a_{2}}\bar{\eta}_{a_{3}}D_{a_{3}}
-\eta_{a_{2}}\bar{\eta}_{a_{1}}D_{a_{1}}]
\end{eqnarray}
and $\beta$ is arbitrary coefficient.
For coupling function $f$, with non-vanishing coefficients in the Taylor expansion to any order (like exponent of \cite{Sagnotti:2010at},\cite{Fotopoulos:2010ay},\cite{Manvelyan:2010je}), this operator generates all terms in the cubic interaction of any three HS fields with any possible number of derivatives $\Delta$ in the range $\Delta_{min}\leq \Delta \leq \Delta_{max}$. The leading term for this interactions is (\ref{leading}). All interactions of HS gauge fields in flat space-time of any dimensions $d\geq 4$ with any number of derivatives are unique and are generated by the generating function (\ref{gf}). It is possible to write this Generating Function in another form, which is equivalent to this one due to partial integration and field redefinition in free Lagrangian \cite{Manvelyan:2010je}.

There is a subset of these cubic interaction vertices, which doesn't mix with other vertices in any order in flat space (assuming there is consistent nonlinear action). These are \emph{minimal cubic selfinteractions} of Higher Spin fields of \cite{Manvelyan:2010wp}. Minimal cubic selfinteraction for higher spin $s$ field includes $s$ derivatives in the cubic interaction, and $s-1$ derivatives in the first order on field gauge transformation. This kind of selfinteraction is a straightforward generalization of Yang-Mills theory and linearized gravity. The generating function (\ref{gf}) (for $\beta=0$) generates only this subset of vertices. Note, that for $\beta=0$, in the first order expansion of (\ref{gf}) (for colored spin one field), gives Yang-Mills cubic vertex, while in the second order - cubic vertex of Einstein-Hilbert gravity (with appropriate field redefinition, discussed in \cite{Manvelyan:2010wp}).

In fact, the exponential form of the generating function (\ref{exp}) in \cite{Manvelyan:2010je} is chosen arbitrarily, motivated by string theoretical analysis of \cite{Sagnotti:2010at}. The relation between cubic interactions of reducible \cite{Fotopoulos:2010ay} and irreducible \cite{Manvelyan:2010je} settings should be studied more carefully.

\subsection*{Acknowledgements}

\quad The author is grateful to E. Joung, D. Francia, R. Manvelyan, R. Mkrtchyan, W. R\"uhl and A. Sagnotti for discussions on the subject of this work.
This work is supported in part by Alexander von Humboldt Foundation under 3.4-Fokoop-ARM/1059429 and CRDF-NFSAT-SCS MES RA ECSP 09\_01/A-31.

\end{document}